\documentclass[prd,twocolumn,nofootinbib]{revtex4}
\pdfoutput=1
\usepackage{amsmath,amssymb,graphicx}
\usepackage{epsfig}

\begin{document}

\preprint{\hepth{0904.2386}}

\title{Global-Local Duality in Eternal Inflation}

\author{Raphael Bousso and I-Sheng Yang\\
  Center for Theoretical Physics, Department of Physics\\
  University of California, Berkeley, CA 94720-7300, U.S.A.\\
  {\em and}\\
  Lawrence Berkeley National Laboratory, Berkeley, CA 94720-8162,
  U.S.A.}

\begin{abstract}
  We prove that the light-cone time cut-off on the multiverse
  defines the same probabilities as a causal patch with initial
  conditions in the longest-lived metastable vacuum.  This establishes
  the complete equivalence of two measures of eternal inflation which
  naively appear very different (though both are motivated by
  holography).  The duality can be traced to an underlying geometric
  relation which we identify.
\end{abstract}

\maketitle

\section{Introduction}
\label{sec-intro}

In an eternally inflating spacetime, anything that is not completely
forbidden will happen infinitely many times.  To define relative
probabilities, various regularization procedures, or ``measures'',
have been explored, including~\cite{LinMez93,LinLin94,
  GarLin94,GarLin94a,GarLin95,
  GarSch05,VanVil06,Van06,Bou06,Pag06b,Lin06,Pag07,Pag08,DGSV08,BouFre08b,
  GarVil08,Win08a,Win08b,Win08c,LinVan08,Pag09,Bou09}.  Some measures are
formulated as geometric cut-offs: The relative probability of events of
type $I$ and $J$ is defined in terms of the ratio of the number of
occurrences of each type of event, $N_I$ and $N_J$, in some finite
portion of the spacetime.

Geometric cut-offs proposed so far can be classified as ``global'' or
``local''.  Global cut-offs define a time slicing in the multiverse and
compute relative probabilities as a late-time limit:
\begin{equation}
  \frac{p_I}{p_J}=\lim_{t\to\infty} \frac{N_I(t)}{N_J(t)}~,
  \label{eq-global}
\end{equation}
where $N_I(t)$ is the number of occurrences prior to the time $t$.  The
result depends strongly on the choice of time foliation, so there are
many inequivalent ways to define probabilities by a global cut-off.

Local cut-offs consider the number of events in a finite neighborhood
of a single inextendible timelike geodesic in the multiverse.
Relative probabilities are defined by the number of occurrences in
this finite neighborhood, averaged over initial conditions and
possible decoherent histories:
\begin{equation}
  \frac{p_I}{p_J}=\frac{\langle N_I(t) \rangle
  }{\langle N_J(t) \rangle }~.
  \label{eq-local}
\end{equation}
The result depends on how the neighborhood is defined, and on the
initial conditions used, so that there are many inequivalent measures
that can be obtained from local cut-offs.  Interestingly, however, both
local prescriptions studied so far~\cite{Bou06,BouFre08b} have a global
``dual''.

The first global-local duality was described in Ref.~\cite{BouFre08b}:
The (global) scale factor time
cut-off~\cite{LinMez93,LinLin94,GarLin94,
  GarLin94a,GarLin95,Lin06,DGSV08,BouFre08b} is dual to the (local)
``fat geodesic'' cut-off, in which the neighborhood of the geodesic is
chosen to have fixed physical volume, and one averages over geodesics
starting in a particular vacuum: that which occupies the greatest
proper volume fraction at late scale factor time.  This duality is
somewhat limited, because the definition of scale factor time is
ambiguous in collapsed regions such as
galaxies~\cite{BouFre08b,DGSV08}.  The scale factor/fat geodesic
duality holds only in universes without collapsed regions, where the
global cut-off is unambiguous.\footnote{Fat geodesics are always
  well-defined, so it is natural to ask if the duality can be made
  more general.  However, it is not clear whether there exists a
  global foliation that reduces to scale factor time in expanding
  regions but reproduces the probabilities computed from fat geodesics
  even in collapsed regions.}

In this paper, we will prove another global-local duality: The
(global) light-cone time cut-off~\cite{Bou09} is dual to the (local)
``causal patch'' cut-off~\cite{Bou06}, in which the relevant
neighborhood of the geodesic $g$ is defined as the causal past $C(g)$
of the entire geodesic.  The duality holds if one averages over causal
patches generated by geodesics starting in a particular vacuum: that
which occupies most horizon volumes at late light-cone time.

Our proof generalizes a much less powerful argument given in
Ref.~\cite{Bou09}, which proceeded by showing that the difference
between relative probabilities computed from two different global
cut-offs (light-cone time and scale factor time) is the same as the
difference between relative probabilities computed from two local
cut-offs (causal patch and fat geodesic).  The known scale factor/fat
geodesic duality~\cite{BouFre08b} then implied the claimed
light-cone/causal patch duality.  Of course, that argument could only
be as general as the scale factor/fat geodesic duality it relied on,
so it applied only in everywhere-expanding universes. Additional
assumptions rendered the argument still less general: it applied only
to multiverse regions that are homogeneous, isotropic, and spatially
flat on the horizon scale.

\begin{figure*}[t]
\begin{center}
\includegraphics[scale = .6]{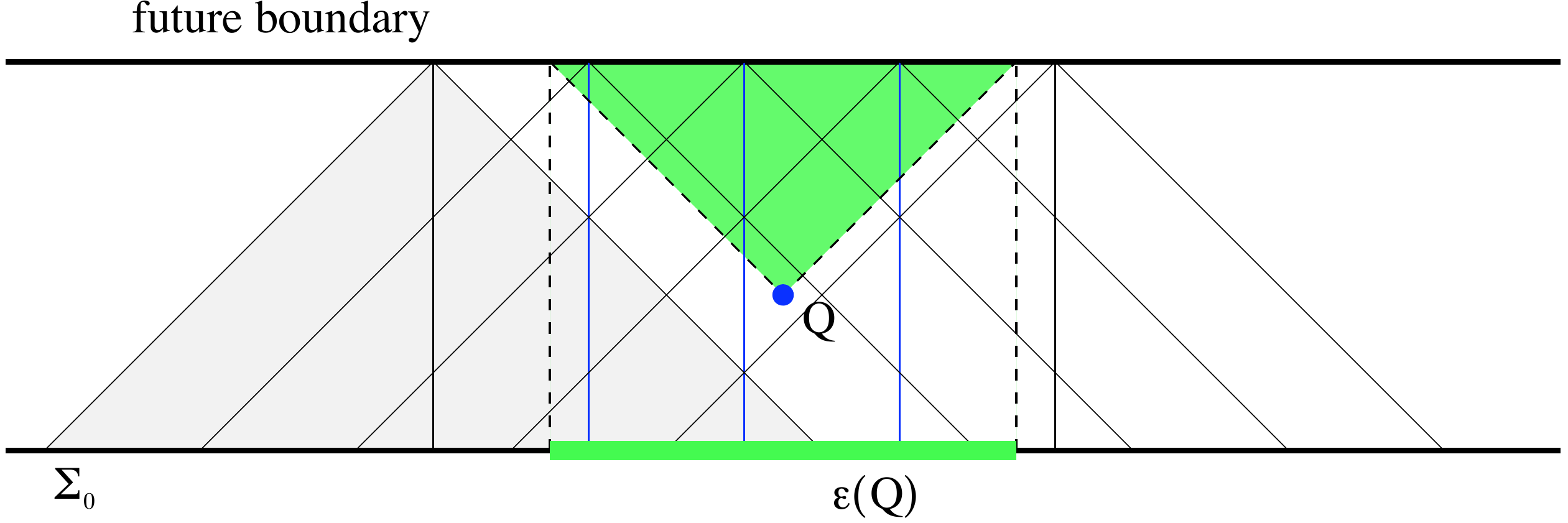}
\end{center}
\caption{Geodesics (thin vertical lines) emanating from an initial
  surface $\Sigma_0$ define an ensemble of causal patches (the
  leftmost is shaded grey/light) with a particular mix of initial
  conditions.  The causal patch measure assigns to the event $Q$ a
  weight proportional to the number of patches that contain $Q$.
  Notice that $Q$ is contained precisely in those causal patches whose
  generating geodesics (blue) enter the causal future of $Q$, $I^+(Q)$
  (shaded green/dark).  In the continuum limit, the weight of $Q$ is
  therefore proportional to the volume, $\epsilon(Q)$, of the
  projection of $I^+(Q)$ onto $\Sigma_0$.  This observation is crucial
  to our proof of equivalence to the light-cone time cut-off.  The
  light-cone time of $Q$ is defined as $t(Q)\equiv -\frac{1}{3} \log
  \epsilon(Q)$.}
\label{fig-magic}
\end{figure*}

Our present proof eliminates all of the above restrictions.  We will
establish the light-cone/causal patch duality directly, without
interposing another, less general global-local duality.  We will
assume only that the universe is eternally inflating.  At the center
of our proof is a simple geometric relation. Let $Q$ be some event in
the multiverse, and let $g$ be a timelike geodesic (which need not
contain the event $Q$).  {\em The causal patch $C(g)$ will contain the
  event $Q$, if and only if the geodesic $g$ enters the causal future
  of $Q$.}  This is shown in Fig.~\ref{fig-magic}.

Our argument proceeds by using the same family of geodesics that
define light-cone time, to also define an ensemble of causal patches.
The light-cone time of the event $Q$ is defined as (minus the log of)
the fraction of geodesics that enter the causal future of $Q$, which
by the above relation is the same as the ensemble-fraction of causal
patches that will contain $Q$.  This implies that the local and the
global cut-off will yield the same relative probability for different
types of events, as long as all events occur at the same light-cone
time.  However, the ensemble-fraction depends on light-cone time,
decreasing exponentially as the geodesics are diluted by the
cosmological expansion.  Thus, the causal patch ensemble will weight
later events exponentially less than the light-cone cut-off.

The two measures will nevertheless agree, if this discrepancy affects
all types of events equally, i.e., if the ratio of the rates at which
events of different types occur is independent of time.  But this is
precisely what happens in the late-time attractor regime of the
light-cone slicing, when $N_I(t)$ grows exponentially with time, with
an $I$-independent coefficient.  Therefore, if we use the attractor
regime to define the initial conditions for the ensemble of causal
patches, both measures will agree.\footnote{It is not necessary to use
  the attractor regime as an initial condition in the global measure,
  since the relative probabilities in Eq.~(\ref{eq-global}) are
  dominated by events occurring at late times in any case.}  In a
generic landscape, the attractor regime is completely dominated by the
longest-lived de~Sitter vacuum, so this amounts to starting all but a
negligible fraction of causal patches in this dominant vacuum.

\paragraph{Outline}

In Sec.~\ref{sec-causalpatch}, we show that a spacelike hypersurface
$\Sigma_0$, together with a family of geodesics puncturing it, defines
an ensemble of causal patches with specific initial conditions.  The
weight of a particular event $Q$, according to the causal patch
measure, has a geometric representation as the volume occupied on
$\Sigma_0$ by those geodesics that end up in the causal future of $Q$.
The causal-patch probability for an event of type $I$ is the sum of
the volumes associated with all events of type $I$ occurring in the
spacetime.  Sec.~\ref{sec-lightcone} contains the proof of the
light-cone/causal patch duality.  The proof uses aspects of the
universal late-time behavior of the light-cone slicing, which are
derived in Sec.~\ref{sec-properties}.

\paragraph{Discussion}

We can prove only that the light-cone time and causal patch cut-offs
yield the same measure, not that they yield the correct measure.  To
identify which, if any, of the extant proposals is correct, one can
proceed in two ways: either phenomenologically (mostly, by
falsification), or by derivation from a fundamental theory for which
there exists independent evidence.

The phenomenological approach has been quite
fruitful~\cite{GarGut06,PogVil06,FelHal06,SchVil06,
  BouFre06b,Vil06b,BouHar07,BouFre07,
  BouYan07,CliShe07,OluSch07,GarVil07,DGSV08,Sch08,DGLNSV08,MerAda08,Fre08,
  BouLei08,BouFre08b,Sal09,BouHal09}.  Measures make predictions, some
of which are robust independently of the details of the landscape of
vacua.  A number of global cut-offs are ruled out because of
predictions that conflict dramatically with
observation~\cite{LinLin96,Gut00a,Gut00b,Gut04,
  Teg05,BouFre07,BouFre06b,Pag06, Pag06b,Lin07,Gut07}. It is
interesting that both the scale factor time cut-off and the light-cone
time cut-off, which have so far\footnote{A potential phenomenological
  problem for both measures is the so-called staggering problem.  In
  the BP model~\cite{BP} of the string landscape (and perhaps more
  generally), the dominant vacuum can only decay to vacua with smaller
  cosmological constant if the resulting cosmological constant is
  negative.  Thus, the dominant vacuum can populate the landscape
  efficiently only by first transitioning to vacua with higher
  cosmological constant.  Such upward jumps are exponentially
  suppressed at least by the difference in horizon entropy of the two
  de~Sitter vacua.  As pointed out in Ref.~\cite{SchVil06} (in the
  context of a different measure in which the same issue arises), this
  can lead to a staggered probability distribution: a few vacua are
  strongly favored over all others.  This would eliminate most of the
  landscape, and thus its ability~\cite{Sak84,BP} to solve the
  cosmological constant problem.  As shown by Schwartz-Perlov and
  collaborators~\cite{Sch06,OluSch07,Sch08}, this problem is absent
  for certain ranges of reasonable model parameters.  It remains to be
  seen whether the string theory landscape falls into this range.
  Similarly, both measures may be dominated by Boltzmann brains, but
  only if the string landscape contains sufficiently long-lived
  vacua~\cite{BouFre06b,BouFre08b,DGLNSV08}.} evaded such problems,
have a local dual.  For example, no natural local dual is known for
the proper time cut-off~\cite{LinLin94,
  GarLin94,GarLin94a,GarLin95,Lin07}, a measure that is ruled out
observationally by the youngness
paradox~\cite{LinLin96,Gut00a,Gut00b,Gut04,Teg05,BouFre07,Lin07,Gut07}.

The second approach---the derivation of a measure from a unified
fundamental theory, say, string theory---is less well developed.
However, there may be general principles that must govern such a
theory, and which we may already discern, and we can apply such
principles to the measure problem.  

We are not aware of any principle supporting the scale factor cut-off
or the fat geodesic cut-off.  Meanwhile, both sides of the duality we
establish here---the light-cone time cut-off and the causal patch
cut-off---are, in different ways, motivated by the holographic
principle.  The necessity of restricting the description of spacetime
to a single causal patch was first discovered by studying the
holographic properties of black holes~\cite{SusTho93}.  The light-cone
time slicing~\cite{Bou09} was constructed in response to the
proposal~\cite{GarVil08} that the holographic UV-IR connection of the
AdS/CFT correspondence should have a multiverse analogue.  Both
cut-offs are defined in terms of null hypersurfaces (the event horizon
of a geodesic defines the causal patch; the future light-cone of a
point defines its light-cone time); and indeed, null hypersurfaces are
essential to a general formulation of the holographic
principle~\cite{CEB1,CEB2,RMP}.

The AdS/CFT analogy is most compelling in eternally inflating vacua
(or more precisely, in eternal domains~\cite{Bou09}).  This suggests
that the light-cone time cut-off (and thus, the causal patch) may not
apply to regions with vanishing or negative cosmological constant.
The analogy also suggests that the global cut-off may not be
sharp\footnote{We are grateful to B.~Freivogel, A.~Guth, and
 A.~Vilenkin for stressing this point to us.}, but should be smeared
on timescales of order $|\Lambda_i|^{-1/2}$, where $\Lambda_i$ is the
cosmological constant of vacuum $i$. 
It is intriguing that uncertainties of this
magnitude appear to provide just enough room for resolving two
phenomenological problems: The cut-offs appear to give too much weight
to vacua with negative cosmological constant~\cite{Sal09}; moreover,
they give rise to divergences in supersymmetric vacua with vanishing
cosmological constant~\cite{BouFre06,MerAda08}, where the horizon
scale diverges.  A refinement of the light-cone time/causal patch
cut-off may be needed for these regions.

These limitations illustrate that one can only get so far by
extrapolation and analogy, or by formulating and falsifying purely
geometric proposals.  Nevertheless, we are encouraged by the recent
confluence of phenomenological and first-principle support for the
light-cone time/causal patch cut-off (or some closely related
prescription).  If this proves to be the right direction, we will have
discovered more than a measure: we will know that in the multiverse,
both the causal patch and the future boundary have special
significance.  We may be approaching a milestone, at which the
phenomenological study of the measure problem begins to yield
constraints on the fundamental description of the landscape and the
multiverse.

\section{The causal patch cut-off}
\label{sec-causalpatch}

\begin{figure}[t!]
\begin{center}
\includegraphics[scale=.3]{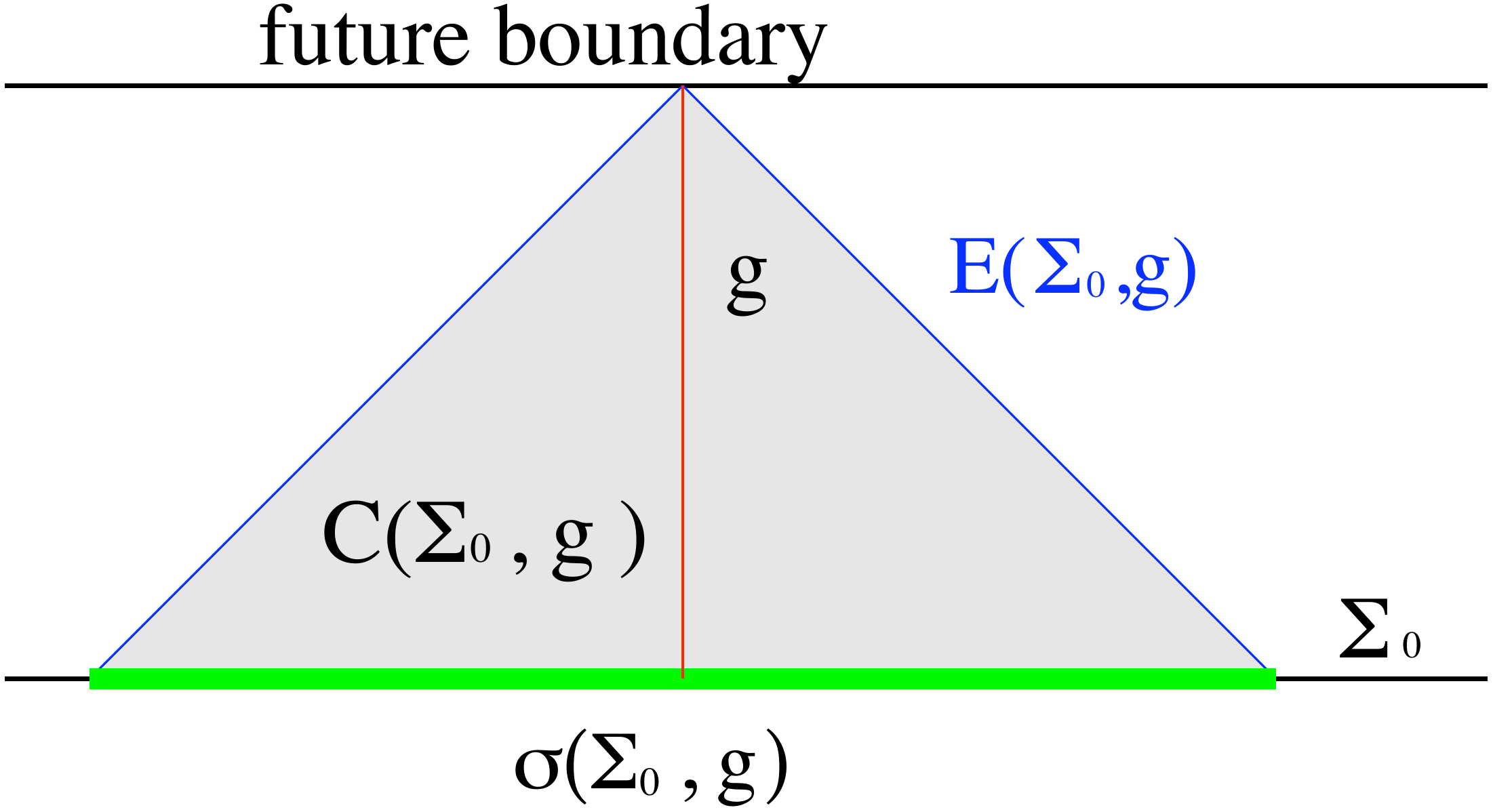}
\caption{\label{fig-causalpatch} A geodesic $g$ starting from an initial   
  surface $\Sigma_0$ defines a causal patch $C$ (shaded region), event 
  horizon, $E$, and initial patch $\sigma$. }
\end{center}
\end{figure}

The causal patch measure assigns to events of type $I$ and $J$ the
relative probability
\begin{equation}
  \frac{\hat P_I}{\hat P_J}=\frac{\langle \hat N_I \rangle
  }{\langle \hat N_J \rangle}~,
\label{eq-ppnn}
\end{equation}
where $\langle \hat N_I \rangle$ is the expectation value of the
number of such events in a particular space-time region: the {\em
  causal patch\/}, defined as the past of an inextendible geodesic $g$
orthogonal to some initial spatial hypersurface $\Sigma_0$:
\begin{equation}
C(\Sigma_0,g)\equiv I^-(g)\cap I^+(\Sigma_0)~.
\end{equation}
That is, the causal patch consists of those points to the future of
$\Sigma_0$ from which some point on $g$ can be reached by a timelike
curve (Fig.~\ref{fig-causalpatch}).  The boundary $\partial C$ of the
causal patch in the spacetime $M$ consists of a null and a spacelike
portion.  The null portion is the {\em event horizon}
\begin{equation}
E(\Sigma_0,g) \equiv \partial C(\Sigma_0,g)\cap I^+(\Sigma_0)~.
\end{equation}
The spacelike portion is the subset of $\Sigma_0$ contained within the
event horizon,
\begin{equation}
\sigma_0(\Sigma_0,g) \equiv \partial C(\Sigma_0,g)\cap \Sigma_0~,
\end{equation}
which we shall call the {\em initial patch\/}.

\subsection{Ensemble of histories and initial conditions}
\label{sec-ensemble}

The appearance of an expectation value, $ \langle \hat N_I \rangle $,
in the above definition indicates that we are considering an ensemble
of causal patches.  Let us take $Z$ identical copies of $\Sigma_0$ and
pick the same starting point for geodesics.  Because of decoherent
quantum effects, the resulting $Z$ causal patches will not be
identical.  For example, the initial vacuum $\alpha$ may decay at
different times and/or into different vacua, etc.~\cite{Bou06}.  Given
initial conditions, the probabilities for different decoherent
histories can be computed as usual from local dynamical laws.  The
expectation value is defined as
\begin{equation}
\langle \hat N_I \rangle=\lim_{Z\to\infty} 
Z^{-1} \sum_{\nu=1}^{Z} N_I(\nu)~,
\label{eq-ensemble1}
\end{equation}
where $N_I(\nu)$ is the number of times the outcome $I$ occurs in the
$\nu$-th causal patch.  We assume that any observations of interest
involve large enough observers or apparatuses that $N_I(\nu)$ (as well
as the space-time geometry) is definite in each decoherent history.
This is certainly true for all observations we make.

In general, the ensemble average, $\langle \hat N_I \rangle$, will
depend on the choice of initial conditions.  A theory of initial
conditions might instruct us to start in one particular initial state
and no other, as was implicitly assumed above.  In general, however,
it may define an ensemble of initial conditions.  For example, it may
tell us to start in the empty metastable de Sitter vacuum $\alpha$
with probability $p^{(0)}_\alpha$, with $\sum_\alpha
p^{(0)}_\alpha=1$.\footnote{We will aim to use lower case variables
  (e.g., $p$) and indices (such as $i,j,\ldots$) when referring to
  vacua.  Greek indices $\alpha, \beta, \ldots$ refer specifically to
  metastable de~Sitter vacua; the longest-lived metastable vacuum is
  called $*$. Indices $m,n,\ldots$ refer to terminal vacua (vacua with
  $\Lambda\leq 0$).  We will use capitalized variables ($N, P,\ldots$)
  and indices ($I,J,\ldots$) to refer to events.}  In this case, we
should enlarge the ensemble of Eq.~(\ref{eq-ensemble1}) and include a
weighted average over of initial conditions.  Eq.~(\ref{eq-ensemble1})
still holds, but instead of constructing all $Z$ causal patches from
the same initial surface $\Sigma_0$, we construct $Z p^{(0)}_\alpha$
patches from an initial surface $\Sigma^\alpha_0$ which is in vacuum
$\alpha$.  More generally, the initial patch could be in a terminal
vacuum, or it may contain matter and radiation or more than one
vacuum; in this case the sum would run over a larger class of possible
initial regions.  Such refinements will not play an important
quantitative role in this paper, assuming only that the initial
conditions have nonzero support in at least one long-lived metastable
vacuum.

If the initial vacuum is a long-lived metastable de Sitter vacuum
$\alpha$, then the size of the initial patch
$\sigma_0(\Sigma^\alpha_0,g_\alpha)$ is essentially independent of the
future evolution (Fig.~\ref{fig-causalpatch}).  Its boundary is given
by the event horizon of the de Sitter space $\alpha$, a sphere of
radius $H_\alpha^{-1}= (\Lambda_\alpha/3)^{-1/2}$.  This holds true
even if the geodesic later enters a vacuum with very small
cosmological constant, like ours.  The area of the event horizon will
become large, but only after the decay.  If the decay happens $h$
Hubble times after $\Sigma_0$, it will change the horizon size on
$\Sigma^\alpha_0$ by an amount of order $\exp(-h)$ relative to the
event horizon of an eternal de Sitter space with cosmological constant
$\Lambda_\alpha$.  For generic metastable vacua, $h$ is typically
exponentially large, so the horizon area on $\Sigma^\alpha_0$ has
radius $H_\alpha^{-1}$ to superexponential accuracy, independently of
future decays.

Because of this property, we may choose $\Sigma^\alpha_0$ in the above
ensemble to be as small as a single horizon volume, or {\em patch of
  type\/} $\alpha$, which we denote as $\mathring\alpha$.
Geometrically, it is defined as a three-dimensional ball of radius
$H_{\alpha}^{-1}$ with Euclidean metric, i.e., as the interior of the
event horizon on a spatially flat slice of de~Sitter space with
cosmological constant $\Lambda_\alpha$.  Its proper spatial volume is
\begin{equation}
v_\alpha=\frac{4\pi}{3}H_\alpha^{-3}~.
\label{eq-v}
\end{equation}
(The flat 3-geometry is chosen for later convenience: The interior of
most horizon regions of metastable vacua on surfaces of constant
light-cone time is indeed flat to great accuracy.)

\subsection{Global representation of the ensemble}
\label{sec-glorep}

\begin{figure}[t!]
\begin{center}
\includegraphics[scale=.27]{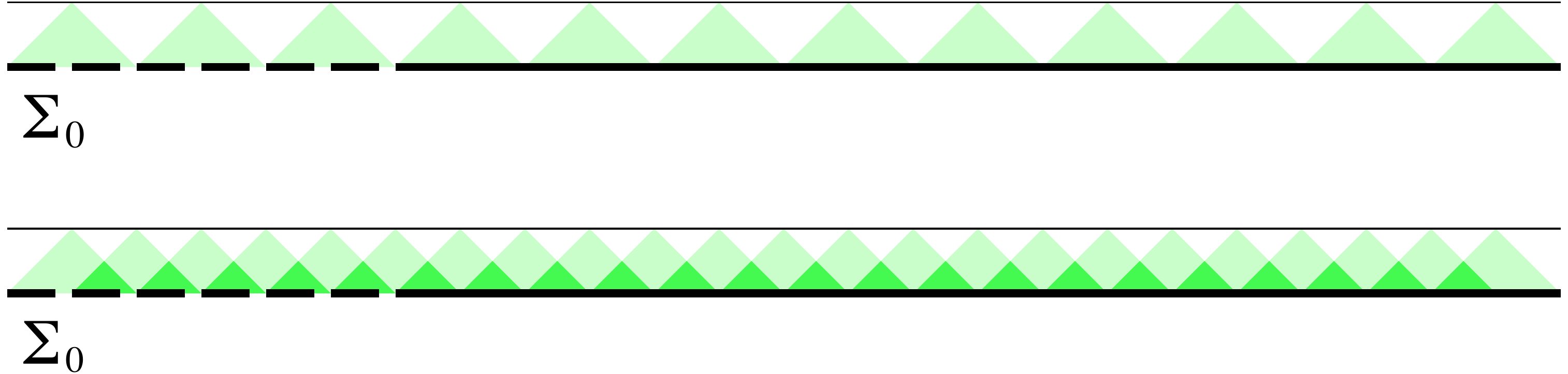}
\caption{Top: An ensemble of causal patches (shaded triangles) can be
  represented in a single large geometry.  Suppose that initial
  conditions require starting in one of two particular de~Sitter
  vacua, with probability $p^{(0)}_1=0.25$ and $p^{(0)}_2=0.75$.  Let
  $\Sigma_0$ be a spacelike hypersurface containing a very large
  number of both types of de~Sitter horizon regions, so that we can
  choose large numbers $Zp^{(0)}_1$ (dashed) and $Zp^{(0)}_2$ (solid)
  of nonoverlapping initial patches.  Then relative
  probabilities for events of type $I$ and $J$ are given directly by
  the ratio $N_I/N_J$ of the numbers of such events in the causal
  patch regions.---Conversely, any $\Sigma_0$ and set of geodesics
  emanating from it defines an ensemble of causal diamonds.
  Increasing the density of geodesics enlarges the ensemble (bottom);
  an event occuring, say, in two different patches counts twice.  If
  each vacuum region contains many horizon volumes, this will not
  change the statistical properties of the ensemble.}
\label{fig-ensemble}
\end{center}
\end{figure}

We have defined probabilities in terms of an ensemble of causal
patches, averaging both over initial conditions and over decoherent
histories.  It is easy to see that one can represent the ensemble of
$Z$ distinct causal patches in a single large geometry, by enlarging
the initial surface $\Sigma_0$ to include $Z$ nonoverlapping horizon
volumes, of which a fraction $p^{(0)}_\alpha$ is in vacuum $\alpha$.
Let us write this schematically as
\begin{equation}
  \Sigma_0\supset \sum_\alpha (Z p^{(0)}_\alpha)~ \mathring\alpha~.
\label{eq-z}
\end{equation}
By constructing one causal patch from each initial patch
$\mathring\alpha$ (Fig.~\ref{fig-ensemble}), one recovers the ensemble
that appears in Eq.~(\ref{eq-ensemble1}).  In this representation,
events $N_I(\nu)$ can be thought of as occurring in the same universe
for different $\nu$ (though they will not all be accessible to the
same observer).

Conversely, we can regard any large initial hypersurface $\Sigma_0$,
along with a set of timelike geodesics originating from $\Sigma_0$, as
defining an ensemble of initial conditions for the causal patch
measure.  For example, let $\Sigma_0$ be spatially flat, containing a
volume $\bar Z p^{(0)}_\alpha v_\alpha$ of each de~Sitter vacuum
$\alpha$, where $\bar Z$ is very large.  The region occupied by vacuum
$\alpha$ need not be connected, but we will assume that each portion
has volume much greater than $v_\alpha$, so that boundary
effects\footnote{Regions of different vacua are separated by
  two-dimensional boundaries.  Near the boundaries, general relativity
  imposes nontrivial constraints on the geometry and extrinsic
  curvature of $\Sigma_0$.  Physically, a boundary will typically
  consist of a domain wall that typically expands into the region of
  higher cosmological constant.} can be neglected.  (This assumption
will be satisfied on the surfaces of constant light-cone time that we
will consider as initial surfaces below.)  In addition to $\Sigma_0$,
we must specify the $Z$ points at which orthogonal geodesics should be
erected, defining $Z$ causal patches.  If we choose these points to
form, say, a rectangular grid, with spacing $2H_\alpha^{-1}$ in
regions of vacuum $\alpha$, then we will have defined an ensemble
consisting of $Z p^{(0)}_\alpha$ nonoverlapping causal patches
starting with vacuum $\alpha$, where $Z/\bar Z$ is a number of order
unity that depends on the grid shape and does not affect relative
probabilities.

Since we have already assumed that boundary effects are not important,
we can be sure that the statistical properties of the ensemble will
not change if we increase the density of geodesics, for example by
including another geodesic midway between any pair of neighboring
starting points on $\Sigma_0$ (Fig.~\ref{fig-ensemble}).  The patches will 
now overlap, and the same event may be counted by more than one patch.  But 
each event will be overcounted by the same factor, so this will not affect 
relative probabilities.  More generally, relative probabilities will be
unchanged as long as the density of geodesics in regions of vacuum
$\alpha$ is given by
\begin{equation} 
  \rho_\gamma(0)=z/v_\alpha~,
\end{equation}
for any $z\geq 1$.  It is convenient to renormalize
Eq.~(\ref{eq-ensemble1}):
\begin{equation}
  \langle \hat N_I \rangle=z^{-1}\lim_{Z\to\infty} 
  Z^{-1} \sum_{\nu=1}^{Zz} N_I(\nu)~.
\label{eq-ensemble2}
\end{equation}
This allows us to take the limit $z\to\infty$ without changing
relative probabilities or encountering divergences.

\subsection{Probability as initial volume}
\label{sec-invol}

The geometric picture we have developed for the ensemble average
allows us to represent the probability for a certain type of event in
terms of volumes on $\Sigma_0$.  Consider a particular event $Q$ of
type $I$, as shown in Fig.~\ref{fig-magic}.  This event will be
included in any causal patch whose generating geodesic $g$ enters the
chronological future of $Q$, $I^+(Q)$.  Therefore, its total
probability is proportional to the number of geodesics entering
$I^+(Q)$.  If we had chosen to place geodesics at a fixed density per
proper volume on $\Sigma_0$, the probability of $Q$ would thus be
proportional to the volume, $\epsilon(Q)$, on $\Sigma_0$, of those
geodesics that enter $I^+(Q)$.  Since we have instead chosen to
consider a fixed number of geodesics per horizon patch
$\mathring\alpha$, the probability of $Q$ is equal to the {\em patch
  number\/} $\pi(Q)$:
\begin{equation}
  \hat P(Q)=(Zz)^{-1}\, \pi(Q)~,
\label{eq-pzp}
\end{equation}
where the patch number is defined as the fraction of a patch, on
$\Sigma_0$, taken up by the starting points of the geodesics that
enter $ I^+(Q)$:
\begin{equation}
\pi(Q)\equiv\frac{\epsilon(Q)}{v_\alpha}~.
\label{eq-patch0}
\end{equation}
In other words, $\pi$ is the volume of the starting points measured
in units of the horizon volume given in Eq.~(\ref{eq-v}).

Because any two non-overlapping horizon patches on $\Sigma_0$ are
likely to remain causally disconnected, their causal patches cannot
both contain $Q$.  Therefore we have $\pi(Q)\lesssim 1$. If $Q$
occurs after many Hubble times of de Sitter expansion, then $\pi(Q)$
will be exponentially small.  Therefore, we can neglect the
probability that the starting points cover more than one vacuum;
indeed, Eq.~(\ref{eq-patch0}) assumes that all geodesics that enter
the future of $Q$ started in the same vacuum.  

Since $\pi(Q)$ is independent of $Z$, any individual event $Q$ will
have vanishing probability in the large $Z$ limit.  But we are
interested in the probability for events of type $I$, not just in one
particular instance of such an event.  In the global picture of
eternal inflation, events of any type will occur infinitely many times
in the future of $\Sigma_0$.  The probability for an event of type
$I$, according the to the global representation of the causal patch
measure we have developed, is the sum of the patch number of each
instance: (Fig.~\ref{fig-magic}):
\begin{equation}
  \hat P_I\propto \sum_{Q\in I} \hat P(Q)~,
\label{eq-vql}
\end{equation}
where the sum is over all events of type $I$ and $\hat P(Q)$ is
defined in Eq.~(\ref{eq-pzp}).  The notation ``$\propto$'' indicates
that an $I$-independent normalization factor has been dropped.  Thus,
Eq.~(\ref{eq-vql}) defines relative probabilities for events of type
$I$ and $J$.

\section{Equivalence to the light-cone time cut-off}
\label{sec-lightcone}

Light-cone time is defined as follows~\cite{Bou09}: Let
$\gamma(\Sigma'_0)$ be the congruence of geodesics orthogonal to the
hypersurface $\Sigma'_0$, and let $Q$ be an event in the future of
$\Sigma'_0$.  The light-cone time $t$ at $Q$ is defined in terms of
the patch number\footnote{In Ref.~\cite{Bou09}, the light-cone time
  was defined in terms of the proper volume of starting points on
  $\Sigma'_0$.  This distinction can be absorbed into a deformation of
  the initial hypersurface.  Because relative probabilities are
  independent of the choice of $\Sigma'_0$, they are in particular
  unaffected by this modification.  The present choice will serve us
  better for formal reasons.} $\pi(Q)$, on $\Sigma'_0$, of the
starting points of those geodesics that enter the future of $Q$,
$I^+(Q)$:
\begin{equation}
t(Q)=-\frac{1}{3}\log \pi(Q)~.
\end{equation}

In the light-cone cut-off measure, the relative probability of events
of type $I$ and type $J$ is defined as the limit
\begin{equation}
  \frac{\check P_I}{\check P_J}=\lim_{t\to\infty}
  \frac{N_I(t)}{N_J(t)}~,
\label{eq-lc}
\end{equation}
where $N_I(t)$ is the number of events $Q_I$ of type $I$ whose
light-cone time is less than $t$.  We will now show that this measure
is equivalent to the causal patch measure defined in the previous
section, with a suitable choice of initial hypersurface $\Sigma_0$.

The main ingredient of this proof is the following assumption: At late
times, the number of events of any type $I$ grows at the same
universal exponential rate,
\begin{equation}
\langle N_I\rangle = \check{N_I} e^{\gamma t}+ O(e^{\varphi t})~,
\label{eq-hacek}
\end{equation}
with $0<\gamma<3$, up to subdominant effects, $\varphi<\gamma$, whose
relative contribution can be neglected at late times.  Moreover, the
number of horizon patches of metastable vacua grows at the same
universal rate:
\begin{equation}
  \langle n_\alpha\rangle = \check n_\alpha e^{\gamma t}+ O(e^{\varphi t})~.
\label{eq-hacek2}
\end{equation}
We will later justify this assumption rigorously and derive the values
of $\check{N_I}$ and $\gamma$ from parameters of the landscape.  For
now, we may take universal exponential growth to be a defining
characteristic of eternal inflation.

By Eqs.~(\ref{eq-lc}) and (\ref{eq-hacek}), the light-cone measure
gives probabilities
\begin{equation}
  \check P_I\propto \check N_I~.
\end{equation}
Let us compare this to the causal patch measure with initial
conditions defined in the manner described in Sec.~\ref{sec-glorep}.
Specifically, we consider the ensemble of patches generated by the
geodesics in $\gamma$, starting from a large hypersurface $\Sigma_0$
which we take to be a surface of constant light-cone time $t_0>0$.

Consider the geodesics that enter the future light-cone of a point
$Q\in\Sigma_0$.  By definition, they occupy a patch number
$\pi=e^{-3t_0}$ on $\Sigma'_0$.  Moreover, if $Q$ lies in a vacuum
de~Sitter region,\footnote{We shall find in the following section that
  this is the case for all but a superexponentially small fraction of
  the volume of $\Sigma_0$, which can be neglected at this stage.
  This does not mean that we will be neglecting regions containing
  matter when we count events.}  then the same geodesics occupy
exactly 1 horizon patch on $\Sigma_0$, and are orthogonal to
$\Sigma_0$.  (This follows, for example, from the arguments given in
Ref.~\cite{Bou09}, which apply in the vacuum limit.)  Therefore, if we
started with $z'$ geodesics per horizon volume on $\Sigma'_0$, there
will be
\begin{equation}
  z= z' \pi(t_0)=z' e^{-3t_0}
\end{equation}
geodesics per horizon volume on $\Sigma_0$.  In particular, the number
of geodesics per horizon volume is constant on $\Sigma_0$, so the
construction summarized in Eq.~(\ref{eq-ensemble2}) can be applied.

Let us choose $t_0$ so large that the correction term in
Eq.~(\ref{eq-hacek2}) can be neglected.  Then the surface
$\Sigma_0$ will satisfy Eq.~(\ref{eq-z}) with $p^{(0)}_\alpha
\propto \check n_\alpha$.  By Eq.~(\ref{eq-hacek2}), increasing $t_0$
any further is equivalent to increasing $Z$ in Eq.~(\ref{eq-z}), so it
leaves relative probabilities untouched.

By Eq.~(\ref{eq-vql}), the causal patch measure defines relative
probabilities
\begin{equation}
  \hat P_I \propto
  \int_{t_0}^{\infty} dt~\frac{d\langle 
    N_I(t) \rangle}{dt}~Z^{-1}\pi(t)~.
\end{equation}
Note that $\pi$ depends only on $t$: Because light-cone time is
defined in terms of patch number, the patch number of an event $Q$
depends only on the light-cone time at which it takes place.
Substituting Eq.~(\ref{eq-hacek}), the integral is trivial,
\begin{equation} 
  \hat P_I \propto  
  \int_{t_0}^{\infty} dt\, \gamma\, \check N_I \, e^{(\gamma-3)t}
  \propto \check N_I\propto \check P_I~,
\end{equation} 
and we find that normalized probabilities are the same as in the
light-cone cut-off measure.  (We remind the reader that ``$\propto$''
signifies equality up to $I$-independent factors, which do not affect
relative probabilities.)

\section{Properties of the light-cone cut-off}
\label{sec-properties}

In this section we will establish a number of key properties of
light-cone time, including the results used in the previous section
for the proof of equivalence to the causal patch measure,
Eqs.~(\ref{eq-hacek}) and (\ref{eq-hacek2}).  We will begin with two
simple examples and then consider the general case.  Since it is clear
from the context which quantities should be thought of as expectation
values, we will omit the brackets $\langle \rangle$ in the interest of
readability.

\subsection{Pure de~Sitter} 

Let us first consider a completely stable vacuum with positive
cosmological constant $3H_*^2$, which we call $*$. Strictly, this case
is outside the scope of this paper, since there are no terminal vacua,
but it provides a useful starting point.  The metric of the
corresponding de~Sitter geometry, in flat coordinates,
is 
\begin{equation} 
  ds^2=-dT^2+H_*^{-2} e^{2H_*T} [dr^2+r^2
  (d\theta^2+\sin^2\theta d\phi^2)]~.
  \label{eq-flatds}
\end{equation}
Let us choose $\Sigma'_0$ to be a finite volume of the hypersurface
$T=0$, with radius $r_0\gg 1$.  The orthogonal congruence $\gamma$
consists of the comoving worldlines at fixed $(r,\theta,\phi)$.  It
follows trivially from the symmetries of this choice that surfaces of
constant $T$ must also be surfaces of constant light-cone time, but it
will be instructive to derive the relation $t(T)$.  Consider a point
$Q$ at time $T$; by homogeneity, we can assume $r=0$ without loss of
generality.  The future light-cone of $Q$ has comoving radius
$e^{-H_*T}$ at future infinity.  The proper volume, on $\Sigma'_0$, of
the geodesics entering this light-cone is
$\epsilon(Q)=\frac{4\pi}{3H_*^3}\exp(-3H_*T)$.  Since the volume of a
single horizon patch is $v=\frac{4\pi}{3H_*^3}$, the patch number is
$\pi(Q)=\exp(-3H_*T)$, and the light-cone time is
$t(Q)=-\frac{1}{3}\log\pi(Q)=H_*T$.  In terms of light-cone time, the
metric is
\begin{equation}
  ds^2=H_*^{-2} \left( -dt^2+e^{2t} 
    [dr^2+r^2 (d\theta^2+\sin^2\theta d\phi^2)]\right)~.
  \label{eq-flatds2}
\end{equation}

It follows that the number of horizon patches is given by
\begin{equation}
n_*(t)=\check n_* \exp(3t)~,
\label{eq-nstable}
\end{equation}
with $\check n_*=r_0^3$. Pure de~Sitter space is in a thermal state,
and events occur at a Boltzmann-suppressed rate per Hubble volume and
Hubble time. Let $\kappa_{I*}$ be the rate at which events of type $I$
(e.g., the formation of a Boltzmann brain) occur. Then
\begin{equation}
N_I=\kappa_{I*} n_*(t)~.
\label{eq-thermal}
\end{equation}
Therefore, Eqs.~(\ref{eq-hacek}) and (\ref{eq-hacek2}) are satisfied
with $\gamma=3$ and $\check N_I=\kappa_{I*} \check n_*$.

\subsection{Single metastable vacuum}

We have claimed that $\gamma<3$; this holds in any landscape that has
terminal vacua, or sinks. To see this, let us now consider the case of
a single metastable de~Sitter vacuum, which we call $*$. It can decay
into terminal vacua by the nucleation of bubbles, at small
dimensionless rates $\kappa_{m*}$ per Hubble volume and Hubble
time. Let us choose the same initial surface as in the previous
example of a stable de~Sitter vacuum. Wherever the vacuum has not
decayed, the metric is described by Eq.~(\ref{eq-flatds}), and the
relation $t=H_*T$ will hold.

Let us find the correction to Eq.~(\ref{eq-nstable}) due to
decays. For small total decay rate $\kappa_*
\equiv\sum_m\kappa_{m*}\ll 1$, we can treat decays as a small
perturbation of the global geometry; that is, we will work at leading
order in $\kappa_*$.  The expected number of nucleation events $dN$
between the time $t$ and $t+dt$ is given by $\frac{4\pi}{3}\kappa_*
H_*^4$ times the enclosed physical four-volume:
\begin{equation}
  \frac{d  N  }{dt}= \kappa_*  n_*(t)  ~.
\end{equation}
Note that we are not distinguishing between decays into different
terminal vacua at this stage.

Let us assume model parameters such that all initial bubble radii are
much smaller than the de~Sitter horizon $H_*^{-1}$. Then the evolution
of a bubble can be approximated by the future light-cone of the
nucleation event. Again, by homogeneity, we can consider a decay at
$r=0$, at time $t_n$. At the time $t$, the bubble will have comoving
radius $r_b(t,t_n)=e^{-t_n}-e^{-t}$. It will have destroyed a physical
volume $\frac{4\pi}{3H_*^3} r_b^3 \exp(3t)$ of the vacuum $*$,
corresponding to
\begin{equation}
  \frac{d  }{dN}\,\delta n_*   = -\left(e^{t-t_n}-1\right)^3 
\end{equation}
lost horizon patches per bubble. (Here we have neglected collisions
between bubbles, which is legitimate at leading order in $\kappa_*$.)
This can be written as $ \frac{d }{dN}\, \delta n_*= e^{3(t-t_n)}
[1-O(e^{-(t-t_n)})]$.

It follows that at late times, $t-t_n\gg 1$, the bubble occupies
precisely the volume that a single horizon patch at $t_n$ would have
expanded to by the time $t$, up to exponentially small
corrections~\cite{GarSch05}. Thus, we will make a negligible error by
assuming that the bubble forms immediately at its asymptotic comoving
size, and treating the bubble wall as comoving. This simplifies the
derivation of the evolution equation for $ n_*(t)  $.
During a time $dt$, the de~Sitter expansion produces $3  n_*(t)
 dt$ new horizon volumes, and $\kappa_* n_*(t) dt$ horizon
patches are lost to decay. Thus, 
\begin{equation}
\frac{d  n_* }{dt}= (3-\kappa_*)  n_*(t)~,
\label{eq-singlerate}
\end{equation}
and it follows that
\begin{equation}
 n_*(t)  =\check n_* e^{(3-\kappa_*)t}~.
\end{equation}
Therefore, Eq.~(\ref{eq-hacek2}) is satisfied with $\gamma=3-\kappa_*$
and $\check n_*=r_0^3$.

The number of terminal bubbles of type $m$ produced between $t$ and
$t+dt$ is
\begin{equation}
  \frac{dN_m}{dt}=\kappa_{m*} n_*(t)~.
\label{eq-bubc}
\end{equation}
At late times, all bubbles of type $m$ are statistically equivalent,
because their production is a local effect in an empty de~Sitter
region. Therefore, the expected number of events of type $I$ per
bubble, $dN_I/dN_m$, will depend only on the type of bubble, and on
the time since bubble nucleation, $\tau\equiv t-t_n$.

To find the total number of events of type $I$ at late times, we
integrate over all types of bubbles and all nucleation times:
\begin{equation}
  N_I(t)=\kappa_{I*} n_*(t)+
  \sum_m \int_0^t
  \left( \frac{dN_I}{dN_m}\right)_{t-t_n}
  \left( \frac{dN_m}{dt}\right)_{t_n}  dt_n~.
\label{eq-krx}
\end{equation}
(The first term is analogous to Eq.~(\ref{eq-thermal}) and takes into
account events that occur in the de~Sitter vacuum.)  Combining the
above equations we find
\begin{equation}
  N_I(t) = \left( \kappa_{I*} + \sum_m N_{Im} \kappa_{m*}\right)n_*(t)~,
\end{equation} 
where
\begin{equation}
  N_{Im}\equiv
  \int_0^\infty d\tau\,  e^{-\gamma\tau} 
  \left( \frac{dN_I}{dN_m}\right)_{\tau}
\label{eq-hrx}
\end{equation}
is independent of time.  Therefore, Eqs.~(\ref{eq-hacek}) is satisfied
with 
\begin{equation}
  \check N_I=( \kappa_{I*} + \sum_m N_{Im} \kappa_{m*})\check n_*~.
\label{eq-evc}
\end{equation}

The upper limit of integration in Eq.~(\ref{eq-hrx}) should strictly
be $t$, so this result is valid only at late times, but this is the
only regime relevant for computing relative probabilities.  For the
measure to be well-defined, the indefinite integral must converge.
This will be the case if $dN_I/dN_m$ diverges nowhere and grows less
rapidly than $e^{\gamma\tau}$ at large $\tau$. If the terminal vacuum
$m$ has negative cosmological constant, then these conditions are
satisfied. Although events can arise with fixed density on infinite
spatially open hypersurfaces inside the bubble, at any finite $\tau$
only a finite portion of every open slice is included, so the integral
is finite for finite $t$. At late times, the size of this portion will
grow no faster than $\exp(2\tau)$. For small $\kappa_*$, this is
slower than $\exp(\gamma\tau)$, so the integral remains finite as
$t\to\infty$. This explains why the ``edges'' of the bubble do not
contribute a divergence. Near the center, the same conclusion follows
from the fact that vacua with negative cosmological constant crunch
after a finite proper time. (Light-cone time is formally infinite at
the singularity, but it will be finite one Planck time before the big
crunch, where the semiclassical description breaks down.)

However, if the vacuum $m$ has vanishing cosmological constant, and if
it contains events of type $I$, then $N_{Im}$ can diverge. In this
case, the light-cone cut-off does not succeed in regularizing the
spacetime. Possible resolutions are discussed in
Ref.~\cite{Bou09}. The potential divergences in $\Lambda=0$ vacua do
not affect our claim of equivalence to the causal patch cut-off, since
the latter would encounter the same
divergence~\cite{BouFre06,MerAda08}. For the purposes of this paper, we
will exclude the interiors of $\Lambda=0$ bubbles (defined more
rigorously as ``hat domains'' in Ref.~\cite{Bou09}). This means we
will be computing relative probabilities for events not occurring in
such regions.

\subsection{General landscape}

Consider a theory such as the string landscape, which contains
metastable de~Sitter vacua $\alpha,\beta,\ldots$ and terminal vacua
$m,n,\ldots$.  We will assume that the metastable vacua are
long-lived, $\kappa_\alpha\ll 1$; states that do not satisfy this
condition can be treated as excited states in the vacua they decay
into.  In this limit, and for the purpose of computing the abundances
of horizon patches of each metastable vacuum, $n_\alpha$, we may
neglect transitory effects such as bubble expansion and the initial
presence of matter and radiation, which affect the size and growth of
de~Sitter regions only in an exponentially small fraction of their
lifetime and volume. The analysis preceding Eq.~(\ref{eq-singlerate})
now yields the rate equation
\begin{equation}
  \frac{dn_{\alpha}}{dt}= (3-\kappa_{\alpha}) n_{\alpha} +\sum_{\beta}
    \kappa_{{\alpha}{\beta}} n_{\beta}~.
\label{eq-rate}
\end{equation}
The first term corresponds to the de~Sitter expansion and to the loss
of horizon patches due to the decay of the vacuum $\alpha$.  The final
sum, which did not appear in the previous subsection, describes the
production of $\alpha$-patches by other metastable vacua $\beta$.

This matrix equation takes exactly the same form\footnote{However, the
  equation is for a different physical variable: In
  Ref.~\cite{GarSch05}, it is for the volume occupied by the vacuum
  $\alpha$; here is is for the number of horizon patches of vacuum
  $\alpha$.  Consequently, the dominant vacuum we find below is
  exactly the same as the vacuum dominating the scale factor
  cut-off. But because of the difference in measures, it dominates in
  a different sense: In the light-cone cut-off, it dominates the
  number of horizon patches, whereas in the scale factor measure it
  dominates the proper volume.  (There is another distinction, which
  is trivial: The term $3 n_\alpha$ on the right hand side is absent
  in Ref.~\cite{GarSch05}, because the volume fractions rather than
  total volume are described.)} as Eq.~(37) in Ref.~\cite{GarSch05},
and it has the same mathematical solution, which takes the form given
in Eq.~(\ref{eq-hacek2}):
\begin{equation}
  n_\alpha(t) = \check n_\alpha e^{\gamma t}+ O(e^{\varphi t})~.
\label{eq-h2}
\end{equation}
Here $\gamma$ is the largest eigenvalue of the matrix
$M_{\alpha\beta}$, and $\check n_\alpha$ is the corresponding
eigenvector; $\varphi$ is the second-largest eigenvalue.  Arguments
given in the appendices of Ref.~\cite{GarSch05} generalize
straightforwardly to show that $\varphi<\gamma<3$.

Since the decay of metastable vacua is an exponentially suppressed
tunneling process, the decay rates will vary enormously, and there is
generically one vacuum with much longer life time than all others.  We
will call this the dominant vacuum, $*$.  A straightforward
generalization of arguments presented in Ref.~\cite{SchVil06} shows
that the above eigenvector is dominated by the $*$ vacuum, and the
associated eigenvalue is related to its total decay rate, $\kappa_*$,
\begin{equation}
  \check n_\alpha\approx 
  \delta_{\alpha *}~,~~~\gamma\approx 3-\kappa_*~,
\end{equation}
to exponentially good approximation.

We conclude that at late times, the number of patches of every vacuum
grows at a universal rate, governed by the decay rate of the
longest-lived metastable vacuum.  Since the growth is exponential,
this asymptotic regime will completely dominate over all earlier
transitory regimes, and we can compute probabilities from it alone.
Therefore, we may as well assume that the initial surface $\Sigma'_0$
is already in the asymptotic regime, allowing us to drop terms of
order $e^{\varphi t}$ and smaller.

To obtain an expression for the number of events of type $I$ and
derive Eq.~(\ref{eq-hacek}), we can now proceed in close analogy with
Eqs.~(\ref{eq-bubc})--(\ref{eq-evc}).  At the time $t$, bubbles of
type $i$ are produced at the rate
\begin{equation}
  \frac{dN_i}{dt}=\sum_{\alpha} \kappa_{i\alpha} n_\alpha(t)~.
\end{equation}
The total number of events of type $I$ is
\begin{eqnarray} 
  N_I(t) & = & \kappa_{I*} n_*(t)+
  \sum_{i\neq *} \int_0^{\!\!\!\!\!\! i~\,t} 
  \left( \frac{dN_I}{dN_i}\right)_{t-t_n}
  \left( \frac{dN_i}{dt}\right)_{t_n}  dt_n
  \label{eq-krx2}\\
  & = &\left( \kappa_{I*}\check n_* + 
  \sum_i \sum_\alpha N_{Ii} \kappa_{i\alpha}\check n_\alpha\right)
  e^{\gamma t}~,
\label{eq-krx3}
\end{eqnarray} 
where
\begin{equation}
  N_{Ii}\equiv
  \int_0^{\!\!\!\!\!\! i~\,\infty} d\tau\,  e^{-\gamma\tau} 
  \left( \frac{dN_I}{dN_i}\right)_{\tau}~.
\end{equation}
To avoid overcounting, the integral should run only over a single
bubble of vacuum $i$, excluding regions of other vacua nucleated
inside the $i$ bubble; this restriction is denoted by index $i$
appearing on the upper left of the integration symbol.  We conclude
that Eq.~(\ref{eq-hacek}) is satisfied with $\check N_I=( \kappa_{I*}
\check n_* + \sum_i \sum_\alpha N_{Ii} \kappa_{i\alpha}\check
n_\alpha)$.

The integral $N_{Im}$ will be finite, and the measure well-defined,
under the condition identified in the previous subsection: the absence
of $\Lambda=0$ vacua or at least of observations therein.  In
particular, there is no divergence associated with the thermal
productions of events at late times in metastable vacua $\alpha$,
since the number of such events in a single bubble grows like the
number of horizon patches, which is by definition slower than the
growth rate $e^{\gamma \tau}$ of the dominant vacuum.

\acknowledgments We are grateful to B.~Freivogel for very helpful
discussions.  This work was supported by the Berkeley Center for
Theoretical Physics, by a CAREER grant (award number 0349351) of the
National Science Foundation, and by the US Department of Energy under
Contract DE-AC02-05CH11231.

\bibliographystyle{apsrev4-1}
\bibliography{all}
\end{document}